\documentclass[usegraphicx,usenatbib,useapjfonts,apj]{emulateapj}
\usepackage{amssymb}
\begin{document}

\title{Model Atmospheres for Irradiated Giant Stars: Implications for the Galactic Center}

\author{Raul Jimenez,\altaffilmark{1} Juliana P. da Silva,\altaffilmark{2,4} S. Peng Oh,\altaffilmark{3} Uffe G.  J{\o}rgensen,\altaffilmark{4} David Merritt\altaffilmark{5}}

\affil{}

\altaffiltext{1}{Department of Physics and Astronomy, University of Pennsylvania, 
Philadelphia, PA 19104, USA; raulj@physics.upenn.edu}
\altaffiltext{2}{Department of Physics, Universidade Federal de Minas Gerais, Brasil; ju.dasilva@gmail.com}
\altaffiltext{3}{Department of Physics, University of California, Santa Barbara, CA 93106, USA; peng@physics.ucsb.edu}
\altaffiltext{4}{Niels Bohr Institute, Copenhagen, DK-2100, Denmark; uffegj@nbi.dk}
\altaffiltext{5}{Department of Physics, Rochester Institute of Technology, Rochester, NY, USA; merritt@astro.rit.edu}

\begin{abstract}
Irradiation of a stellar atmosphere by an external source (e.g. an AGN) changes its structure and therefore its spectrum. 
Using a state-of-the-art stellar atmosphere code, 
we calculate the infrared spectra of such irradiated and transformed stars. 
We show that the original spectrum of the star, 
which is dominated by molecular bands,
changes dramatically when irradiated even by a low-luminosity AGN
($L_{\rm X}  = 10^{33}$ erg s$^{-1}$), becoming dominated by atomic lines 
in absorption.
We study the changes in the spectrum of  low-mass carbon- and oxygen-rich 
giant stars as they are irradiated by a modest AGN,
similar to the one at the Galactic center (GC). 
The resulting spectra are similar to those of the faintest S-cluster 
stars observed in the GC.  
The spectrum of a star irradiated by a much brighter AGN,
like that powered by a tidally disrupted star, is very 
different from that of any star currently observed near the GC. For the first time we have discovered that the structure of the atmosphere of an irradiated giant changes dramatically and induces a double inversion layer. 
We show that irradiation at the current level can explain the 
observed trend of CO band intensities decreasing as a function of increasing proximity to Sg $A^{*}$. 
This may indicate that (contrary to previous claims) there is no paucity of old giants in the GC, which coexist
simultaneously with young massive stars.
\end{abstract}


\section{Introduction}

The fact that UV radiation and X-rays can alter the atmospheres of stars
has been recognized for more than thirty years 
\citep{DavidsonOstriker73,BaskoSunyaev73,Arons73,Basko+74,Fabian79}.
Irradiation by a source like an active galactic nucleus (AGN) will produce an
increase in the atmospheric temperature and in the mass loss rate
\citep{Edwards80,VoitShull88,ChiuDraine98}. 
Even a modest level of irradiation from a low-luminosity AGN,
like the one currently at the center of the Milky Way, 
can be sufficient to destroy molecules formed in the atmosphere of 
cool giant stars, thus transforming their spectrum without inducing
significant mass loss. 
Recently, \citet{Barman+2004} carried out detailed computations 
of the atmospheric structure of an M dwarf irradiated by a hot 
stellar companion (a pre-cataclysmic variable). 
However, up to now, and despite recent advances in the calculation 
of molecular opacities \citep{Jorgensen05} and in stellar modeling
algorithms,
no detailed computations of the atmosphere of a cool giant star
that is irradiated by an external source have been performed. 

In this paper, for the first time, we compute the stellar spectrum and atmospheric structure 
of a cool ($\sim 4000$ K) giant (both carbon and oxygen rich) star 
in the presence of an AGN. 
Our stellar atmosphere code includes a complete frequency dependent description of the 
atomic and molecular lines that dominate the infrared (IR) spectrum.
Our computational approach is general, but we focus here on
the spectrum of a star that is irradiated by a source at the 
Galactic center (GC).
We are motivated by the recent discovery \citep{Revnivtsev+04} 
that the GC may have been a low-luminosity AGN
($L\approx 10^{39}$ erg s$^{-1}$)
as recently as a couple of hundred years ago. 
In addition, recent estimates of the rate of stellar tidal 
disruptions by the GC supermassive black hole (SBH) 
\citep{merritt04,merritt05}
suggest a rate of order one event per $10^4$ yr or higher
for solar-mass stars.
Tidal disruption of a star by a SBH is expected to produce an extremely
luminous event, $L\approx 10^{44}$ erg s$^{-1}$, 
with a duration of weeks or months (e.g. \citet{Komossa02}).

The GC SBH was recently found to be
surrounded by a cluster of apparently young stars. 
For one of these stars, S0-2, an IR spectrum showed no CO 
absorption, and possible HeI ($2.1$ $\mu$m) absorption \citep{Ghez+2003}. 
The latter indicates that the star cannot have an effective 
temperature less than about $15,000$ K \citep{Hanson+1996}. 
The orbit of S0-2 has a pericenter distance of $\sim 100$ AU 
and an apocenter distance of $\sim 1000$ AU. 
The presence of the HeI line and the absence of CO point to S0-2 being young, 
with spectral class in the domain O and B. 
The case for hot stars so close to the GC has been further reinforced 
by \citet{Eisenhauer+05} who obtained high S/N spectra of 17 S-cluster stars. 
For the brightest stars in this sample, spectra clearly show the presence 
of the HeI line at $2.1127 \mu$m. 
However for stars with $K > 15$ the line is not present. 
Hence, the spectral properties of the 
S-cluster stars are not uniform.   
If these stars are actually young (a few Myr), their formation so
close to GC SBH is a serious theoretical challenge \citep{Phinney89}.
One possibility is that these stars are not young but old, 
and that their atmospheres have been modified by some physical mechanism: 
stellar collisions, tidal stripping or external radiation (e.g.  \citet{genzel+03,alexandermorris03,HM03}, see also the review by \citet{alexander05} and references therein).

In this work, we compute the detailed 
spectrum of giant stars irradiated by an AGN in the luminosity range 
$10^{33} - 10^{44}$ erg s$^{-1}$ integrated over the  range $2-200$ keV (hereafter we denote the luminosity in this range by $L_{\rm X}$.  In particular, we show how these old giants 
are affected by low-luminosity AGN ($L_{\rm X} \approx 10^{39}$erg s$^{-1}$)
like the one that may have been present at the GC in the recent past, 
and also by the high luminosity ($L_{\rm X} \approx 10^{44}$ erg s$^{-1}$) that is believed
to accompany the tidal disruption of a star by the SBH. 
We find that the spectrum of  an irradiated old giant star is similar to
that of the faintest S-cluster stars ($K > 15$) observed by 
\citet{Eisenhauer+05}. 
However, we are unable to transform the spectra of cool giants 
such that they resemble those of the brightest S-cluster stars 
for a low luminosity AGN ($ L_{\rm X} < 10^{39}$ erg ${\rm s^{-1}}$). Higher luminosities though, as the one from a tidally disrupted star are sufficient to heat up the stellar atmosphere above $15,000$ K, and therefore 
produce the HeI line.  
However, the total destruction of molecular lines in the spectra 
of these giants seems to point out that the deficiency of giants 
in the GC \citep{Sellgren+1990} might not be such. 
We show that, within our model, irradiation explains the decreasing 
CO band intensity observed as a function of radius from Sgr A$^{*}$.
  The other main result of our study is that we find a double layer inversion in the 
 atmospheres of irradiated giants for irradiation fluxes $f > 10^{2}$ erg s$^{-1}$ cm$^{-2}$ on 
 the surface of the star. 

We demonstrate that even if the AGN was shining at or near the 
Eddington luminosity for $10^6$ years, due to the high eccentricity 
of the S-cluster stars orbits, the amount of irradiation is most likely not 
sufficient to make the HeI line appear. 
On the other hand, stars on more circular orbits could potentially be sufficiently transformed to reproduce the spectral properties of  
the brightest S-cluster stars (including the HeI line). 
  
\section{Irradiation of stellar atmospheres: the IR spectrum}

If the atmosphere of the star is
sufficiently irradiated, the energy deposited in the atmosphere will heat
up the outer layers and produce a wind.

\citet{BaskoSunyaev73} constructed a semi-analytic model to account
for the effects of irradiation on a stellar atmosphere. The most
obvious consequence of irradiation is that the temperature of the
area of the star facing the AGN increases by $T_{\rm s}/T_i\sim
(1+F_x/F_i)^{(1/4)}$. At the microscopic level the X-rays photo-ionize
He I, He II and O, C, Ne, N, Fe and its ions. Below $5000$ K the main
source of opacity is due to photo-detachment of $H^{-}$, while above
this temperature opacity is dominated by photo-ionization of oxygen and
carbon.  \citet{BaskoSunyaev73} find that in their
models the envelope develops a significant wind, although most of the
energy is re-radiated. By integrating the hydrodynamic equations, \citet{VoitShull88} 
calculated the rate at which mass is lost from the envelope of red super-giants. 
We use their work to estimate the mass loss rate from giant atmospheres in \S\ref{section:mass_loss}. 

However, in order to compare with the observations of \citet{Eisenhauer+05}, 
we are interested  in computing detailed spectra of irradiated stars. 
For this purpose we have used a stellar atmosphere code \citet{Jorgensen+92}, which is based on the MARCS code \citep{Gustafsson+75}. The models are computed in hydrostatic equilibrium, with
radiative and convective energy transport included. Plane parallel and spherical geometry are considered where appropriate.
The radiative transfer includes neutral and one time ionized atomic lines from the VALD data base and molecular opacities from CO, C2, CN, CS, HCN, C2H2, C3, SiO, TiO, H2O, and several diatomic hydrates (\citet{Jorgensen03,Jorgensen05} and references therein).
All opacities are treated by the opacity sampling technique \citep{HellingJorgensen98}. 
The atmospheric structure and the spectra are  computed separately (in order to allow studies of the
contribution of various species to the spectra individually), but consistently and based on the same linelists.
A new feature of the version of the code used for the present paper is the treatment of external illumination, which we have based on the inclusion
of an improved version of the subroutines developed and described by \citet{AVN99,NV90,VN85}.

For reference, we first compute  the spectra of
non-irradiated oxygen-rich (C/O$=0.0$, T$_{\rm eff}=4000$ K, 
log g=2.0, $Z=Z_{\odot}$)  and carbon-rich (C/O$> 1$) stars. We then consider an irradiated star.  
The irradiation source is taken to be the GC with an AGN spectral shape as given
by \citet{Sazonov+04}.
In particular, we use their eq. 14 for the energy range 
$1 {\rm eV} < E < 2 \, {\rm keV}$ and their eq. 8 for $E > 2$ keV and 23 for $E < 1$ eV.  
The total flux for Sgr A$^*$, assuming a mass of $3.7 \times 10^6$ 
M$_{\odot}$ is $L = 5.0 \times 10^{39} (f_{\rm EDD}/10^{-4})$ erg s$^{-1}$ \citep{Ghez+05}. Although the 
\citet{Sazonov+04} spectral energy distribution is for typical QSO's we use it here as a good approximation to describe the AGN at the GC. Our results are not sensitive to moderate changes on the parameters that describe  the spectral energy distribution of QSO's in \citet{Sazonov+04}.  
For $f_{\rm EDD}=10^{-4}$ the  luminosity of the AGN at the GC corresponds roughly to the estimate by \citet{Revnivtsev+04} for the luminosity of the GC a few hundred years ago. 

\begin{figure} 
\includegraphics[width=\columnwidth]{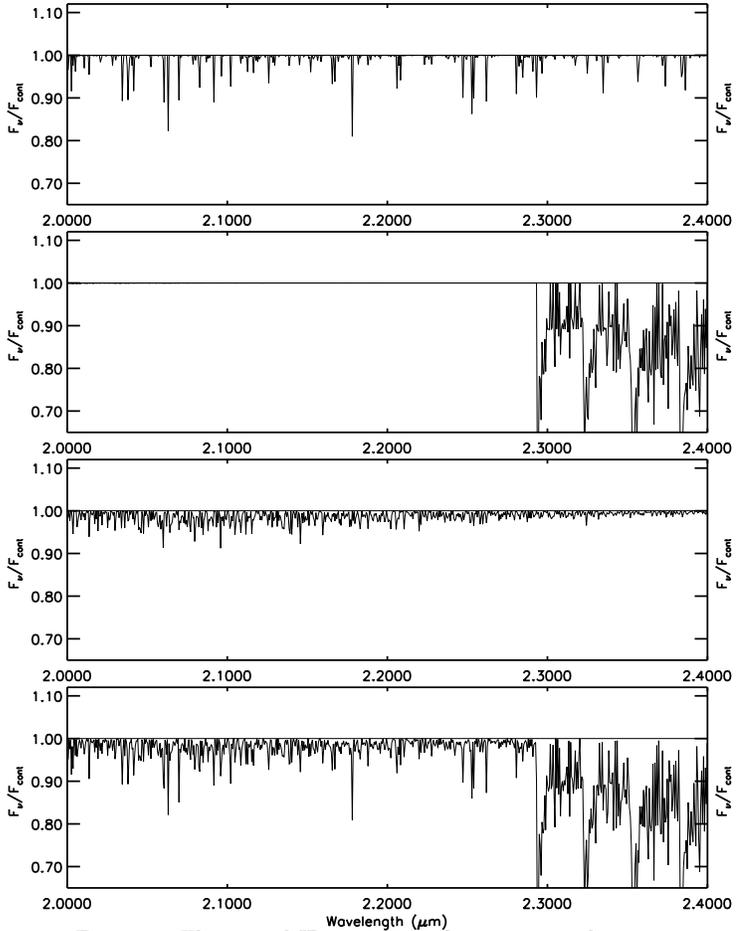}
\caption{Theoretical IR spectrum of an oxygen-rich star at an effective temperature of $4000$K. The upper panel shows the atomic lines, the second panel the CO lines, the third panel other molecules while the bottom panel shows the total spectrum. Note that the strong CO bands start at about $2.3 \mu$, usually beyond the range observed by \citet{Ghez+2003} and \citet{Eisenhauer+05}. 
Note also the presence in the total spectrum of molecular-band 
absorption lines but also of the Br$\gamma$ atomic line at $2.1661 \mu$. }
\label{fig:O} 
\end{figure}

\begin{figure} 
\includegraphics[width=\columnwidth]{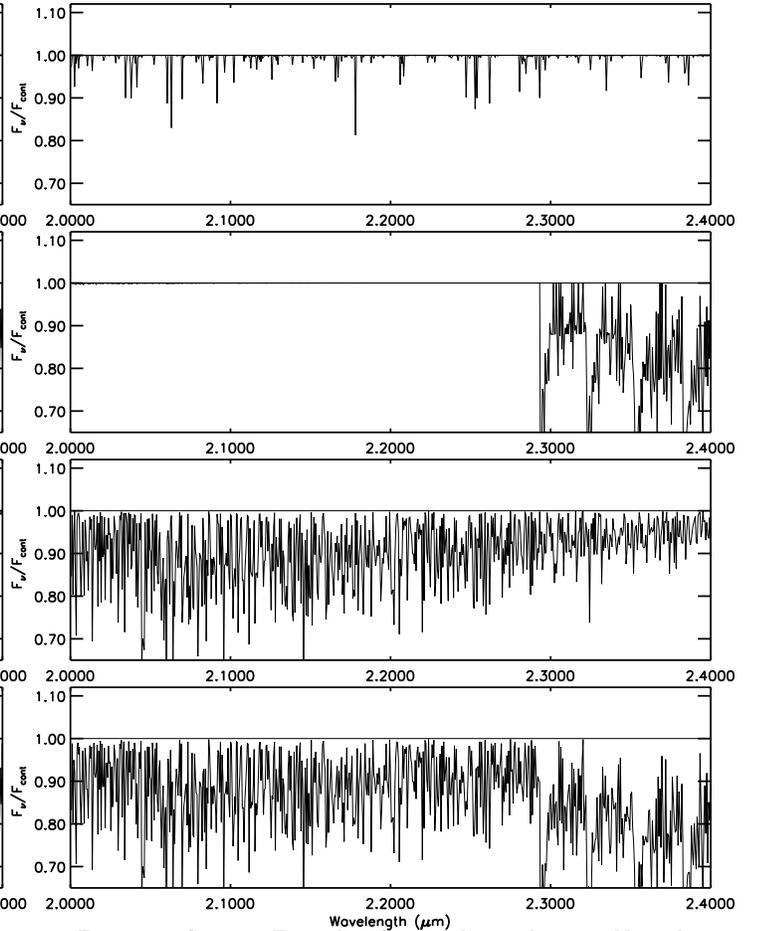}
\caption{ Same as Fig.~\ref{fig:O} but for a carbon-rich giant. 
Note that CO lines are only present beyond $2.3 \mu$, 
while the other molecules get considerably stronger than for an oxygen-rich star.}
\label{fig:C} 
\end{figure}

The IR spectrum ($2 - 2.4 \mu$) for the non-irradiated case is shown 
in Fig.~\ref{fig:O} for the oxygen-rich model. The top panel
shows the atomic lines, the second panel the CO bands,
the third panel other molecular bands and the bottom
panel is the total spectrum. 
Clearly, the IR spectrum of an oxygen-rich
giant is dominated mainly by molecular bands but also shows some atomic lines. The most prominent of these atomic lines is the Br$\gamma$ line at $2.1661 \mu$. 
Note that there are no emission lines. 
It is also worth mentioning that there are no CO lines in the
$2-2.3 \mu$ range; they appear only at wavelengths beyond $2.3 \mu$. 
Therefore, these strong bands would not be observed in the spectra of
\citet{Ghez+2003} or \citet{Eisenhauer+05}.  While the molecular bands are somewhat 
stronger than those observed in the faintest S-cluster stars, the spectrum of a non-irradiated oxygen-rich giant is not too dissimilar from the ones observed for the  faint 
stars in the S-cluster sample (see Fig.~1 in \citet{Ghez+2003} and Fig.~5 in \citet{Eisenhauer+05}).
Fig.~\ref{fig:C} shows the non-irradiated spectrum for a carbon-rich star.  
Note the absence of CO lines below $2.3 \mu$ (they are very strong CO bands beyond this wavelength)  and the increase in the strength of the other molecular bands. Clearly, the non-irradiated spectrum of a C-rich giant does not resemble at all any of the observed S-cluster stars.

\begin{figure} 
\includegraphics[width=\columnwidth]{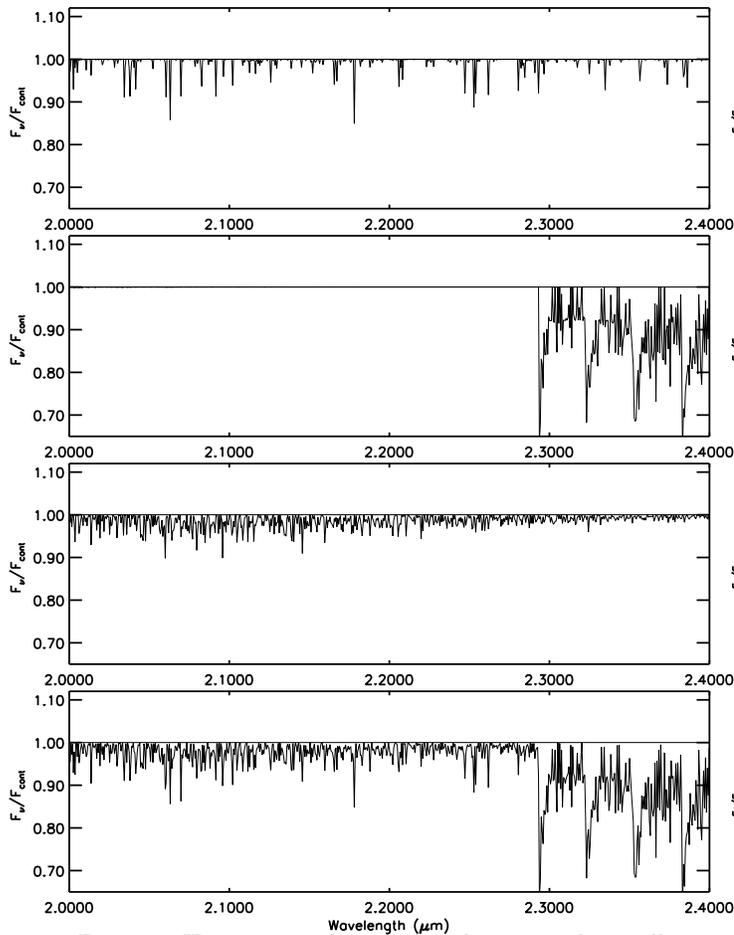}
\caption{IR spectrum of an irradiated oxygen-rich star illuminated with a flux $f = f_o$.
The upper panel shows the atomic lines, the second panel the CO lines, the third panel other molecules while the bottom panel shows the total spectrum. Note the decrease in the strength of the molecular lines with respect to the non-irradiated spectrum of Fig.~\ref{fig:O}. }
\label{fig:Cirr} 
\end{figure}

\begin{figure} 
\includegraphics[width=\columnwidth]{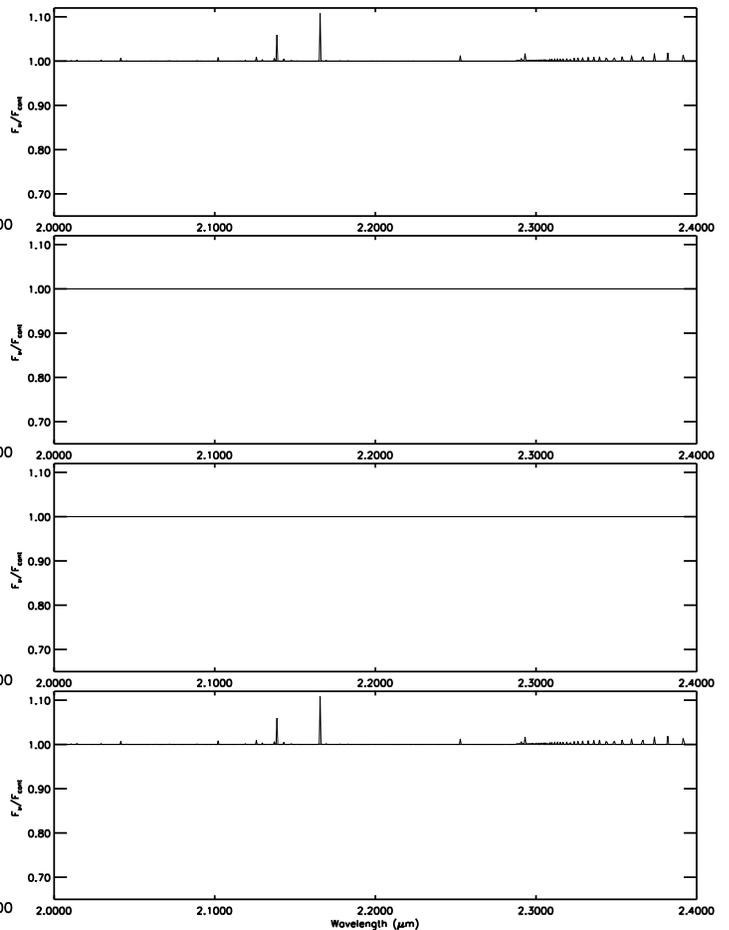}
\caption{Same as Fig.~\ref{fig:Cirr} but for a star with an illumination flux $f = 100f_o$. 
Note the extreme transformation of the spectrum with now only atomic lines in emission and no molecules, including CO.}
\label{fig:Cirr2} 
\end{figure}

\begin{figure} 
\includegraphics[width=\columnwidth]{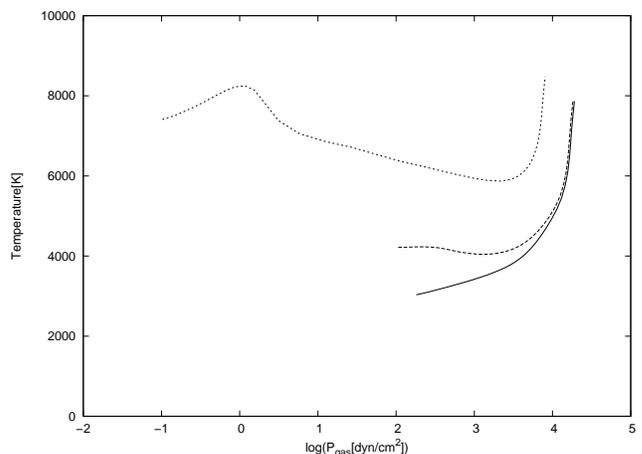}
\caption{Plot of the $T-P_{\rm gas}$ structure of the relevant model atmospheres: solid line is for $f=0$, dashed line is for $f=f_o$ and dotted line is for $f=100f_o$. Note the inversion temperature layer that takes places at illumination fluxes of $f = 100 f_o$}
\label{fig:tpg} 
\end{figure}

\begin{figure} 
\includegraphics[width=\columnwidth]{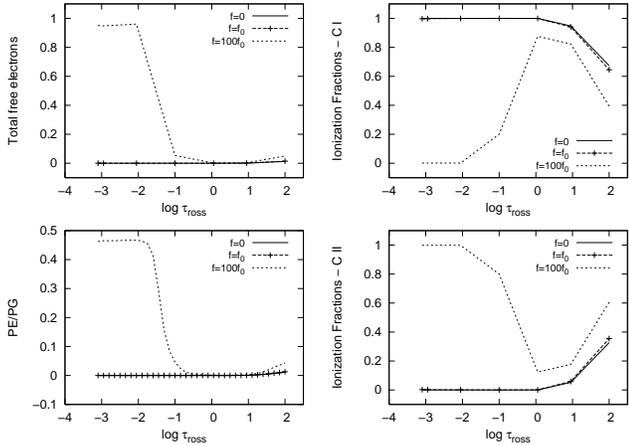}
\caption{The plot shows the value of the ionization fractions of CI and CII (right panels) and electron pressure over gas pressure (bottom-left panel) and total density of free electrons (upper left panel) for non-irradiated and irradiated models.}
\label{fig:intopac} 
\end{figure}

We then irradiate the star as described above, assuming different incident fluxes.
Fig.~\ref{fig:Cirr} assumes an orbit averaged flux:
\begin{equation}
f_{o} = 2 \left( \frac{L}{10^{33} {\rm erg \, s^{-1}}} \right) \left( \frac{f_{\rm supp}} {0.6} \right) \left( \frac{r_{\rm min}}{100 \, {\rm A.U.}} \right) \frac{{\rm erg}}{\rm{ s\, cm^{2}}}
\end{equation} 
where $r_{\rm min}$ is the pericenter distance. See Eq. \ref{eqn:f_suppress} and Fig. \ref{fig:f_suppress} for a definition of $f_{\rm sup}$. On the other hand, Fig.~\ref{fig:Cirr2} assumes a flux of $f= 10^{2} f_{o}$, corresponding to a star illuminated by Sg $A^{*}$ during a more active state when $L\approx 10^{35} {\rm erg s^{-1}}$, still smaller than the luminosity estimated by \citet{Revnivtsev+04} 
for the luminosity of the GC a few hundred years ago. Our illuminated atmospheric models are static and we are not able to compute models with larger illumination fluxes that $f = 10^2 f_o$. For this dynamic models are required, which we are currently constructing. However, even for $f = 10^2 f_o$ the transformation of the spectrum is significant.
 
The first thing to note from Fig.~\ref{fig:Cirr} is the decrease in the strength of the molecular lines even for $f=f_{o}$. Note also the reduction of the CO band intensity. 
If we examine the spectrum of the star for $f=10^{2} f_{o}$, (Fig.~\ref{fig:Cirr2}),
we notice even more significant changes.
As expected, all the molecular bands are gone, including the CO bands. 
However, now some of the atomic lines are in emission due to the stronger irradiation. 
In particular, the Br$\gamma$ line at 2.1661 $\mu$ is now in emission.  
At this small distance, irradiation results in a rise in the temperature 
of the atmosphere at 
$\tau_{\rm ROSS}=10^{-4.9}$ from $2500$ K to $8000$ K. 
However, the HeI line at 2.11 $\mu$ is clearly not present, and more importantly, no other line in the irradiated star appears at the same wavelength.

It is clear that S-cluster stars irradiated at their present orbits ($5 \times 10^{-4}$ to $5 \times 10^{-3}$ pc) by a low-luminosity AGN a few hundred years ago do not resemble the spectra of any of the observed S-cluster stars seen today \citep{Ghez+2003,Eisenhauer+05}. Their spectra would be totally dominated by emission lines. 
On the other hand, if we look at Fig.~\ref{fig:Cirr}, which is equivalent to a star irradiated at a distance of a few hundred AU but with a luminosity of $L_{\rm X}  \sim 10^{33}$ erg s$^{-1}$ the similarity with the faintest stars of the \citet{Eisenhauer+05} sample 
is striking.  In this case, the HeI line is absent and the deepest absorption feature in the 
spectrum is dominated by the Br$\gamma$ line. 

It is clear that after the level of irradiation suggested by \citet{Revnivtsev+04} has ceased, the star will cool down and readjust to its previous 
equilibrium situation in a few years (as soon as the temperature is 
low enough, about $2000$K, molecules will form immediately on a time scale of seconds). However, the 
presence of illumination will stop convection on the  atmosphere. It takes of the order of a few hundred years for 
convection to be restored and therefore the temperature in the outer layers to decrease enough for molecules to be allow to reform. This argument also assures that the time-scale of molecule formation is longer than the rotation time-scale of the stars  assuring that molecules will be wiped out over the whole surface of the star.
 Note however that the present $X$-ray flux from the GC is 
sufficient to destroy molecules, as shown in Fig.~\ref{fig:Cirr}.

Figure 5 shows the temperature versus gas pressure
model structure for a oxygen-rich giant for $f=0,f_{o}$ and $10^{2} f_{o}$.
In the moderately irradiated model $f=f_{o}$ from the GC
(as well as in the non-irradiated model), the upper layers are relatively dense ($P_{\rm gas} \approx$ 100 dyn/cm$^2$).
Therefore the absorption of radiation from the SBH at the GC is substantial already in the top of the atmosphere. As a result, the model structure resembles a photospheric-chromospheric atmosphere 
with a slowly rising, almost flat, chromospheric
temperature distribution. The inner part of the atmosphere is almost unaffected by the illumination at $f=f_{o}$. For $f = 10^2 f_{o}$ the  radiation is strong enough that the atmosphere is heated at all optical depths in the atmospheric model,
and the chromospheric temperature rise is substantial. As a result, the degree of ionization increases, making the continuous opacity increase, whereby the atmosphere expands (considerably).
Figure 6 shows the relative fraction of neutral (C{\sc i}) and
one time ionized (C{\sc ii}) carbon (the two right panels) for the three models in Fig.~5. Other atoms, including H, N, O, Al, Si, S, Ca, and Ni, behave qualitatively similarly (whereas Ca, Mg, Cr and Fe are substantially double
ionized in the top-layers, and He is neutral throughout the
atmosphere). 
The main contributer of free electrons is hydrogen,
and the total abundance and pressure of electrons are shown
in the two left panels. It is seen that the degree of ionization (and the abundance of free electrons) increases rapidly outward from $\log \tau_{\rm ROSS} \approx$ 0 to $-$2.
This is the region of temperature rise in the strongly
irradiated model ($\log {\rm P_{gas}}$ from $\approx 3$ to $0$).
From $\log \tau_{\rm ROSS} = -2$ and outward hydrogen is fully ionized,
and the electron density and pressure therefore now again decreases outward. As a consequence the opacity decreases in the outermost layers, and the
energy deposition due to external illumination also decreases outwards from log $P_{\rm gas}=0$ onwards. The temperature therefore decreases toward the surface from this point onwards,
just like in a normal photospheric model. 
This feature is not seen in any chromospheric model heated from below. It is particular to strongly irradiated atmospheres.

\section{Mass Loss from AGN irradiation}
\label{section:mass_loss}

In this section, we consider whether the heat input from AGN irradiation is sufficient to evaporate the stellar envelope of a star, causing an observable change in its spectrum. 
We consider much larger AGN luminosities than in the previous section,  
up to $\sim 0.5\% L_{\rm Edd}$.
This assumption is motivated by the very high luminosities believed
to accompany stellar tidal disruptions \citep{Komossa02},
and the estimated high rate of such events \citep{merritt04,merritt05},
one per $10^3 -10^4$ yr or so.

This is a radiative rather than a gravitational (e.g., \citet{DaviesKing05}) 
mechanism for stripping stellar envelopes off stars near the GC.
If the effective temperature is only raised to 
about $T_{\rm eff} \approx 15 \,000$ K (which is high enough to produce the observed HeI line), then these modified stars could explained the observed S-cluster stars, though stars with higher effective temperatures will have lifetimes which are too short to account for the number of stars observed (see eq. ~2 in \citet{GP05}). Note that stripping the envelope is a different mechanism than the one proposed in the previous sections of this work where the temperature of the stellar atmosphere was raised due to illumination without mass loss. It is this constant illumination that keeps the stellar atmosphere hot. In the stripping scenario there is no source of heating and the star will cool down to a new equilibrium configuration. 

Regardless, it is still interesting to explore the consequences of 
the mass loss as it will be a useful marker of the past activity of 
Sgr $A^{*}$. 
This mechanism will clearly predict a characteristic dependence of 
the number of "hot" stars on pericentric radius. 
Stripping the envelope off a star exposes its hotter core, 
and thus increases its effective temperature. 
However, the luminosity of the star will be unaffected, 
since the conditions in the stellar core are effectively decoupled 
from the conditions in the envelope. 
With numerical stellar models \citep{JMDPP04}, 
we compute the effective temperature $T_{\rm eff}$ of the stripped star, 
and its new radius $R_{*}$, assuming $L=4 \pi R_{*}^{2} \sigma T_{\rm eff}^{4}$=const. 

Can the entire envelope be stripped due to a close encounter with an AGN? 
\citet{VoitShull88} consider the $X$-ray induced mass loss from stars near 
AGN. 
They consider mass loss due to two mechanisms: thermal winds driven by X-ray heating, and stellar ablation by radiation pressure. For thermally driven winds, they directly integrate the hydrodynamic equations, and find that the formula:
\begin{equation} 
\dot{M}_{\rm thermal}=5.0 \times 10^{-6} \, {\rm M_{\odot} \, yr^{-1}} L_{42}^{0.9} R_{d,15}^{-1.8} R_{*,100}^{2}
\label{eqn:Mdot_thermal}
\end{equation}
reproduces their results very well, as well as the previous analytic results of \citet{Basko+77}. 
Here $L_{42} \equiv L/(10^{42} {\rm erg \, s^{-1}})$ is the AGN luminosity, $R_{d,18}\equiv R_{d}/10^{18} {\rm cm}$ is the distance of the star from the AGN, and $R_{*,100} \equiv R_{*}/100 \, R_{\odot}$ is the stellar radius. Note that this calculation assumes that emission line cooling is quenched in the wind, and is therefore potentially an overestimate. They find that ablative mass loss (which is independent of emission line cooling), is
\begin{equation}
\dot{M}_{\rm abl} \approx 3.8 \times 10^{-6} \, {\rm M_{\odot} \, yr^{-1}} L_{42} R_{d, 15}^{-2} M_{*}^{-1/2} R_{*,100}^{5/2}
\end{equation}
for $R < R_{\rm abl}$, where
\begin{equation}
R_{\rm abl} = 6.3 \times 10^{15} \ {\rm cm} \ M_{*}^{1/2} L_{42}^{1/2} R_{100}.
\label{eqn:Mdot_ablation}
\end{equation}
We will approximate the suppression for $R>R_{\rm abl}$ via $\dot{M}_{\rm abl} \rightarrow \dot{M}_{\rm abl} {\rm exp} (-R/R_{\rm abl})^{2}$. Since the orbital distance $R_{d,15}$, the stellar mass $M_{*}$ and the stellar radius $R_{*,100}$ are all time-dependent, we find the total mass loss by integrating equations (\ref{eqn:Mdot_thermal}) or (\ref{eqn:Mdot_ablation}) numerically. For the AGN luminosity assumed, $L_{42} \sim 5 \times 10^{-3} L_{\rm Edd}$ for a $3 \times 10^{6} M_{\odot}$ SBH, the stellar envelope will be stripped from a star after $\sim 10\%$ of its main sequence lifetime, or $\sim 10^{5}$ years, if the star remains at $\sim 100$ A.U. from the SBH throughout this time. 

Proper motion observations of GC stars have managed to pin down their orbital parameters: specifically, their eccentricity $e$ and pericenter distance $r_{\rm min}$ \citep{Schodel+03,Eisenhauer+05}. The orbits are all highly eccentric; typically $e\approx 0.8-0.9$, while pericenter distances are of order $\sim 100-1000$ A.U. Given these parameters, we can solve for the orbit $R(t)$ implicitly:
\begin{eqnarray}
r &=& \frac{r_{\rm min}(1+e)}{(1+e \, {\rm cos}\theta)} \\ \nonumber
\frac{2\pi t}{\tau} &=& \psi - e \, {\rm sin}\psi \\ \nonumber
{\rm tan}\left(\frac{\theta}{2} \right) &=& {\rm tan} \left[  \left( \frac{1+e}{1-e} \right)^{1/2} {\rm tan} \frac{\psi}{2} \right]
\end{eqnarray}
where $\tau=\left[ 4 \pi^{2} a^{3}/(G M_{\rm AGN})\right]^{1/2}$ is the Keplerian orbital period, and the semi-major axis $a=r_{\rm min}/(1-e)$. In Table 1, we show the results for a $1 M_{\odot}$ and $2 M_{\odot}$ giant undergoing mass loss for $\sim 10^{6}$ years, assuming the observed orbital parameters.

\begin{figure} 
\includegraphics[width=8.5cm]{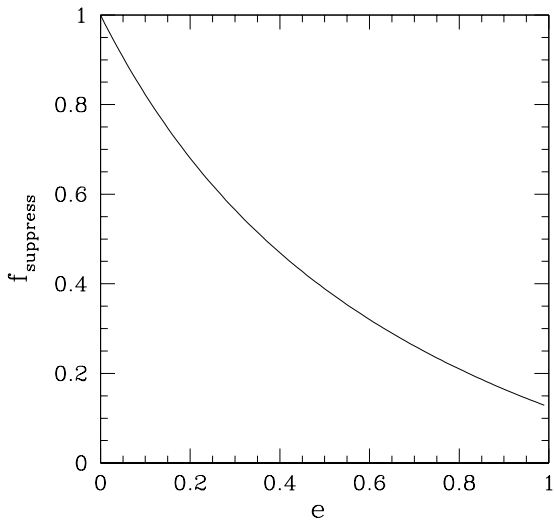}
\caption{The flux weighted fraction of time an object spends close to pericenter, as given by equation (\ref{eqn:f_suppress}), as a function of the eccentricity $e$. This gives the relative reduction in flux compared to a circular orbit at pericenter distance.}
\label{fig:f_suppress} 
\end{figure}

The mass loss can be a significant amount of the star's mass, but is still insufficient to boost the effective temperatures to sufficiently high values. This is because the orbits are highly eccentric, and spend most of their time at large radii, far from pericenter $r_{\rm min}$. We can compute the suppression factor compared to a purely circular orbit by considering the flux-weighted fraction of time an object spends close to pericenter in a single orbit:
\begin{equation}
f_{\rm supp}(e)= \frac{r_{\rm min}^{2}}{\tau} \int_{0}^{\tau}  \frac{dt}{R^{2}(e,t)}. 
\label{eqn:f_suppress}
\end{equation}
(it is acceptable to average over a single orbit since the typical orbital period $\tau \sim 100$ years is much less than the main sequence lifetime; we are estimating the cumulative mass loss over many orbits). This is shown in Fig \ref{fig:f_suppress}. 

\begin{deluxetable}{lcccccc}
\tablecaption{Mass loss and stellar effective temperatures for measured orbits. The orbital parameters are from the fits of Eisenhauer et al (2005). We assume the AGN shines with a luminosity $L=10^{43} {\rm erg \, s^{-1}}$ for $\sim 10^{6}$yrs. The first set of columns is for a 1$M_{\odot}$ star with an initial $T_{\rm eff}=3000$K, $R_{*,100}=1$, the second set of columns is for a 2$M_{\odot}$ star with an initial $T_{\rm eff}=4000$K, $R_{*,100}=1.23$.}
 \tablehead{&  $r_{\rm min}$/AU & $e$ &  $\Delta M$/M$_\odot$ & $T_{\rm eff}$/K & $\Delta M$/M$_\odot$ & $T_{\rm eff}$/K}
\startdata
S2 & 120 & 0.87 &0.31 & 6400 & 0.47 & 7900 \\
S12 & 220 & 0.73 & 0.18 & 5300 & 0.22 & 6300 \\
S14 & 100 & 0.97 & 0.29 & 6200 & 0.51 & 8100 \\ 
S1 & 2020 & 0.62 & 0.35& 3900 & 0.07 & 4900 \\ 
S8 & 180 & 0.98 & 0.15& 5100 & 0.30 & 6900 \\ 
S13 & 1000 & 0.47 & 0.09 & 4400 & 0.14 & 5600 \\ 
\enddata
\end{deluxetable}

\section{The Lack of Giants near the Galactic Center}

\begin{figure} 
\includegraphics[width=\columnwidth]{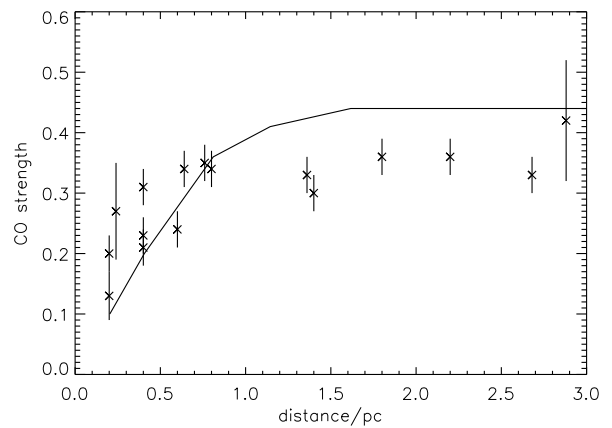}
\caption{Predicted strength of CO (solid line) as a function of distance 
from the GC for models irradiated by an AGN of luminosity $L_{\rm X}  = 10^{33}$ erg s$^{-1}$. Overplotted are CO measurements from 
\citet{Sellgren+1990}.}
\label{fig:co} 
\end{figure}

Given the strong transformation in the atmosphere of old giant stars due
to irradiation, it is worth exploring how the CO abundance correlates with distance from the GC.  
CO observations have been obtained for a dozen stars from
0.2 pc up to 3.6 pc from the GC \citep{Sellgren+1990}.  
They show a clear decrease in the strength of the CO band at 
distances of about 0.5 pc from the GC.
From our numerical experiments we can measure the strength of the CO
absorption as a function of distance from the GC. 
To do this we have irradiated the star with the same 
parameters as in \S 2 at different distances for the luminosity of the AGN at the GC 
today ($L_{\rm X}  = 10^{33}$ erg s$^{-1}$). As can be seen from Fig.~\ref{fig:Cirr}, the irradiated star with $f = f_o$ still contains CO.

Fig.~\ref{fig:co}
shows our prediction (solid line) and the \citet{Sellgren+1990}
data. To compute our predictions we have chosen an average value of the eccentricity $e= 0.77$ from Table 1 and applied the corresponding suppression ($0.2$) factor from Fig.~\ref{fig:f_suppress} to the irradiated flux today.
Although our model is not a perfect fit, the agreement is good and
the general trend is reproduced, namely, a  decrease in CO
absorption band strength the closer the star is to the GC. 
This indicates that
AGN irradiation is producing the right flux of photons to start CO
destruction at a distance of $\sim 1$pc. 
This implies that there might not be a lack of giants 
near the GC and that the only thing we might be seeing is a transformation of the spectrum of the star due to irradiation by the low-luminosity AGN.

\section{Discussion and Conclusions}

It has been argued in the literature (e.g. \citet{Ghez+2003}) that the observed spectra of the S-cluster stars, are in agreement with standard spectra of type B8 or earlier, indicating that
the stars are young, which is a puzzle because at such distances the tidal force by the central BH is far to great to be overcome by densities in normal molecular clouds. However, effects of the radiation field due to Sg $A^{*}$ on stellar atmospheres has hitherto not been taken into account. The upper layers of stars at the distance of the S-cluster will be strongly
affected by this irradiation. We have therefore computed fully self-consistent stellar atmospheres where this irradiation is take into account. The result is a substantial heating of the upper atmosphere. The heating of the upper layers of the atmosphere reduce the intensity of the CO bands as well as all other molecular bands, hereby making even stars
of quite late type look fairly much like the observed S-cluster stars.

In particular, in the spectra from our model atmospheres of irradiated giant stars with T$_{\rm eff} \approx$ 4000K, the intensity of the CO bands is
decreasing when the distance to Sgr A is decreased, in qualitative agreement
with the observations by \citet{Sellgren+1990}. This suggests that contrary to previous claims, there is no dearth of old giants near the galactic center; their molecular signatures have simply been wiped out by the radiation field from Sg$A^{*}$. 
However, some of the observed S-cluster stars have a strong HeI line in their spectra. We have not been able to reproduce this line for irradiated low-mass stars  for realistic values of the GC luminosity, suggesting that that some other mechanism (perhaps recent star formation of massive star) is responsible for their presence.

Our illuminated atmospheric code is static. In fact we have only been able to obtain converged models for values of $f < 100 f_o$. We have not been able with the present static code to predict the structure and spectrum of a star illuminated with $f > 100 f_o$. For doing this a dynamic model is needed. It is not inconceivable that when dynamics effects are included and models are converged for $f > 100 f_o$, the spectra of these irradiated stars will look even more extreme than the models presented in this work. In particular it will be interesting to investigate if the HeI line can be obtained at higher illuminations for dynamical models.    

\section*{acknowledgments}

The research of RJ is partially supported by NSF grants AST-0408698, 
AST-0420920, AST-0437519 and PIRE-0507768  
and by NASA grant NNG05GG01G. 
SPO is supported by NSF AST-0407084. DM is supported by grants AST-0420920 from the NSF and NNG04GJ48G from NASA. JPS acknowledges support from ITA.

%\bibliographystyle{./apj}
%\bibliography{./raul}

\end{document}